\documentclass[useAMS,usenatbib]{mn2e}

\input{tcilatex}

\begin{document}

\title[Carbon in the pregalactic epoch and the search for the first haloes]{%
Carbon in the pregalactic epoch and the search for the first haloes}
\author[A. Lipovka, J. Saucedo and N. Lipovka]{A. Lipovka$^{1}$\thanks{%
E-mail: aal@cajeme.cifus.uson.mx}, J. Saucedo$^{1}$ and N. Lipovka$^{2}$
\\
$^{1}$ Centro de Investigaci\'on en F\'isica, UNISON, Rosales y Blvd.
Transversal, Col. Centro, Edif. 3-I, Hermosillo, Sonora, 83000, M\'exico\\
$^{2}$ Special Astrophysical Observatory RAS, Pulkovskoye Sh. 65, St.
Petersburg, 196140, Russia}
\date{Accepted 2007 .............. . Received 2007 ................ ; in
original form 2007 ............ }
\maketitle

\begin{abstract}
A possible way of detecting first structure formation in a non-standard BBN
Universe during the dark age, due to resonant scattering of the CMBR photons
in the rotational lines of the primordial CH molecule is discussed. The
calculations are made within the framework of the $\Lambda$CDM cosmology and
plausible first structure physical conditions. The carbon chemistry during
the pregalactic epoch is considered. The relative abundance of the CH
molecule is found to be 10(-14) whereas the adopted [C]/[H] ratio was taken
to be 10(-10). The optical depth, dT/T and integration times are estimated.
The calculated optical depth turns out to have high values, that argue in
favor of this molecule as an excellent candidate in searches for first
structure formation. Possible observations with the GMT and ALMA are
discussed.
\end{abstract}

\pagerange{\pageref{firstpage}--\pageref{lastpage}} \pubyear{2007}

\label{firstpage}


\section{Introduction}

The investigation of the epoch of first structure formation in the early
Universe is one of the most intriguing aspects of modern cosmology. The
observational data on the physical conditions during the growth of the first
structures, if detected, would give us extremely important information on
the fundamental problems of the early Universe, such as inflation, Big Bang
Nucleosynthesis (BBN) models (see for example the review by \citet{b30} and
references therein), the formation of the first mini-galaxies (see %
\citet{b36}; for a review, as well as \citet{b22} and reference therein),
first miniquasars \citep{b12} and PopIII stars \citep{b28,b33,b37}. As has
been discussed in previous works, the most promising way to investigate this
epoch, is to search for the Spectral Spatial Fluctuations (SSF), produced by
the primordial molecules in the spectra of the CMBR through resonant
scattering of the CMBR photons. Recently, the possibility of detecting these
lines with ODIN and Herschel satellites has been investigated by \citet{b21}%
. They have also discussed the current observational situation for molecules
based on the light elements. For further details, see \citet{b20}, and the
most recent calculations of the SSF due to the $HD$ molecule in \citet{b24}.
It must be stressed here that in the case of deviation from the simple
equilibrium evolution of the Universe, the abundance of light elements in
some regions of the Universe can increase (see for example possible
production of deuterium by cascades from energetic antiprotons considered in
the paper of \citet{b38}), that can lead to the observability of such
regions.

The SSF from molecules based on the light elements (such as $H$, $D$, $He$, $%
Li$) in the primordial gas has been discussed by many authors during the
last decade. But, perhaps light elements are not the only species capable of
leaving footprints in radio. In recent years, models of Non-standard Big
Bang Nucleosynthesis (NBBN) and new physics have attracted considerable
attention, motivated in part for the discrepancies that remain between the
abundance measurements of $^{4}He$ and $^{2}H$ \citep{b30}. The NBBN models
predict a wide range of abundance for the heavy elements, which could be
used to discriminate, combined with future observational results, among the
alternative models. In this regard, and for reasons that will be discussed
in this paper, the abundance of primordial carbon appears to be of
particular interest. One obvious reason, is related to the peak in abundance
for the $C$ $N$ $O$ group, whose height has a strong dependence on the NBBN
model. In particular, the $^{12}C$ relative abundance can vary from $3\cdot
10^{-14}$ for the standard BBN (see for example \citet{b14}), up to 10$^{-9}$
\citep{b27}. Recently the model of Inhomogeneous BBN has been revisited by %
\citet{b16}, where the relative abundance of $^{12}C$ has been found to be
of the order of $10^{-12}$ \citep{b15}. Besides, the Nonequilibrium
Cosmological Nucleosynthesis discussed in the papers of \citet{b39} and %
\citet{b40} can still further increase its abundance.

The chemistry of the early Universe has also been a subject of investigation
in the last few years. For a review on this topic one may read \citet{b5}
and more recently, the results for the deuterium chemistry in \citet{b6}.
But little effort has been done so far to consider the assumption of the
NBBN, for which, the chemistry of heavy elements has to be included.

Recently, the chemistry of the primordial carbon and oxygen in the early
Universe was calculated \citep{b18} and the possibility of direct
observations of the $CH$ molecules formed in the first haloes was discussed %
\citep{b19,b2}.

It needs to be stressed that not just primordial carbon should be taken into
consideration, but also carbon produced by primeval SN. First stars (PopIII)
possibly appear at $Z\approx 20$, and must significantly enrich the
environment with heavy elements. For example, \citet{b4}, consider a model
in which the relative abundance of $^{12}C$ at $Z=19$ due to feedback can
reach a value of $[C]/[H]=10^{-4}$.

It should also be noted, that the formation epoch of the first stars can be
pushed back to even higher redshifts due to the presence of
ultra-high-energy cosmic rays which stimulated the formation of the first
stellar objects, as it has been recently suggested by \citet{b34}. If this
is the case, one can consider significant abundance of carbon at very early
epochs.

It is for these reasons, that the search for primordial carbon is of special
interest. On the one hand, the detection of primordial CH would help us
discriminate among the NBBN models. On the other, it would allow us to
obtain important information about the formation of first structures in the
Universe (such as minigalaxies, first stars, etc.).

The $CH$ molecule is an excellent species for this aim due to several
reasons:

1)\qquad $^{12}C$ is rather sensitive to the NBBN model.

2)\qquad It is argued here that $C$ stands a better chance for primordial
chemistry than the other two species in the $C$ $N$ $O$ abundance peak of
the BBN element distribution. Although $N$ has an abundance a little bit
higher than $C$, its negative ion $N^{-}$ has a binding energy an order of
magnitude lower than that of $C^{-}$, and the ratios for $NH$ molecule
formation are quite low. With respect to oxygen, its relative abundance is
much lower compared with that of $C$ (see \citet{b14}) and the molecules
based on primordial oxygen are less interesting than those of $CH$ (see also %
\citet{b18}).

3)\qquad $CH$ has a rather high dissociation energy (3.46 eV) and a large
dipole moment.

4)\qquad Because of the rather low densities in the cosmological gas there
is no depletion chain $C-CH -CH_{2}-CH_{3}-CH_{4}$, like the one that exists
in the interstellar medium. For this reason, the kinetics of the ISM is
significantly different to the cosmological one.

Several facilities being planned or under construction will cover the
submillimeter and millimeter wavelength ranges with high sensitivity.
Examples of these are ALMA, CARMA, and the Large Millimeter Telescope (LMT),
under construction in Mexico, whose radiometers will cover the range from
0.8 to 3mm. The redshifted rotational lines of the primordial CH molecule
are expected to fall in the mm region, and the optical depths for first
structures are estimated to be rather high. The feasibility of detecting CH
in the near future, is one of the main motivations for doing detailed
calculations for this molecule.

The aim of the present work is to draw attention to the problem of heavy
elements in the early Universe, and to emphasize on the extremely important
role primordial carbon can play in the investigation of the pregalactic
epoch, as well as in the BBN models. Here we present for the first time, the
complete chemistry of primordial carbon and calculate the expected
footprints of first structure under formation in the rotational lines of the
CH molecule on the basis of the $\Lambda $ - Cold Dark Matter ($\Lambda $
CDM) scenario of cosmic structure formation.

The paper is organized as follows: in the second section we present the
chemistry of primordial carbon in the early universe. In the third, the
observational parameters for first haloes in the dark age epoch are
introduced. Finally, the main conclusions are discussed in the fourth
section.

\bigskip

\section{Chemistry of primordial carbon}

\bigskip

The chemistry of the light elements under the standard BBN model has been
discussed in several papers (see for instance, the review paper \citet{b5})
and will not be discussed in any detail here. Recently the chemistry of
fluorine in the early Universe was investigated by \citet{b26} (see also
references therein for the standard light element chemistry).

The chemistry of the primordial carbon and oxygen in the early Universe was
considered for the first time in \citet{b18}. In the present work the
primordial carbon chemistry is revisited to make it more complete and
up-to-date with more recent data for the molecular processes. Within the
framework of the $\Lambda $ CDM Universe, the parameters adopted for the
calculations are as follows: $\Omega _{m}=0.27$, $\Omega _{b}=0.04$, $\Omega
_{\Lambda }=0.73$ and $H_{0}=71$ km s$^{-1}$Mpc$^{-1}$. (e.g., \citet{b31}.
Standard BBN yields of the light elements are assumed for simplicity, since
their deviation from the standard values do not significantly affect the
carbon chemistry. The kinetic equation of the system written in terms of the
redshift $z$ is given by the following expression: 
\begin{eqnarray}
\frac{dx_{i}}{dz} &=&\frac{1}{H_{0}\left( 1+z\right) \sqrt{\Omega
_{m}(1+z)^{3}+\Omega _{\Lambda }}}\times  \nonumber \\
&&\left[ n_{0}\left( 1+z\right) ^{3}\left(
\sum_{k,l}x_{k}x_{l}R_{kli}-x_{i}\sum_{m,n}x_{m}R_{imn}\right) \right. 
\nonumber \\
&&\left. +\sum_{k}x_{k}R_{ki}-x_{i}\sum_{m}R_{im}\right] ,  \label{Qs}
\end{eqnarray}

where $dz=dtH_{0}\left( 1+z\right) \sqrt{\Omega _{m}(1+z)^{3}+\Omega
_{\Lambda }}$), $n_{0}$ is the density of the gas in the present epoch, $%
x_{i}$ are the dimensionless relative abundance for the $i$ species ( $%
x_{i}=n_{i}/n_{tot}$), $R_{ijk}(T_{c})$ are the rates of the collisional
processes $i+j\rightarrow k$ , as functions of the kinetic temperature $T_{c}
$, $R_{ji}(T_{r})$ are the rates of the radiative processes (formation and
destruction of the molecules by the CMBR photons characterized by the
radiative temperature $T_{r}$). To calculate the kinetic temperature $T_{c}$
we need to solve the following equation:

\begin{eqnarray}
\frac{dT_{c}}{dt} &=&-2T_{c}H_{0}\sqrt{\Omega _{M,0}(1+z)^{3}+\Omega
_{\Lambda }}+  \nonumber \\
&&\frac{2\left( \Gamma _{mol}-\Lambda _{mol}\right) }{3nk}+\frac{\Theta _{ch}%
}{3nk}+\frac{2T_{c}}{3n}\left( \frac{dn}{dt}\right) _{ch}+  \nonumber \\
&&\frac{8\sigma _{t}a_{b}T_{r}^{4}}{3m_{e}c}x_{e}\left( T_{r}-T_{c}\right) .
\end{eqnarray}

Here, the first right-hand term is for the Hubble expansion of the Universe,
the second one corresponds to heating ( $\Gamma _{mol}$ ), or cooling ( $%
\Lambda _{mol}$ ) of the gas by absorption or emission, the third term
describes the heating (cooling) of the medium in exothermic (endothermic)
chemical reactions, the forth one appears here to take into account the
change of the primordial gas density due to chemical reactions which do not
conserve the number of initial species. The last term is due to Thomson
scattering. These equations solve the problem of molecular kinetics at the
pregalactic epoch.

Let us now consider the formation (destruction) processes in more detail.
For the present calculations, whenever appropriate we simply adopt the rate
constant coefficients listed in our previous paper \citep{b18}. But a few
important reactions were missed there, and for others, we have found more
appropriate values for the rate constants. For this reason, in these
calculations we include some new processes and adopt a new set of ratio
constants.

\subsection{Photodetachment}

As it is well known for the case of hydrogen chemistry at low redshifts, the
formation of the molecular hydrogen through the negative ion $H^{-}$ is an
important mechanism. For this reason it is of particular interest to follow
the formation - destruction processes for the negative carbon ion $C^{-}$.
This is a rather stable ion, with a binding energy $D=1.25eV$ (for
comparison, the binding energy of $H^{-}$ is $0.75eV$). It is destroyed by
the radiative photodetachment process:

\begin{equation}
C^{-}+\gamma =C+e^{-}
\end{equation}

The rate for this reaction has been computed from the corresponding
crossection published by \citet{b23} and it is in excellent agreement with
experimental data. The threshold energy for this process is 1.25 eV, and the
rate coefficient is given by the following expression:

\begin{equation}
R_{pd}=3\cdot 10^{5}\frac{T_{r}}{300}\exp (\frac{-15000}{T_{r}}), [s^{-1}]
\end{equation}

which is used in our calculations.

\subsection{Radiative attachment}

\bigskip This is the main channel for the C- negative ion formation and it
is given by the reaction:

\begin{equation}
C+e^{-}=C^{-}+\gamma
\end{equation}

The crossection and the rate coefficient for this process were calculated by %
\citet{b11}, but these authors were interested only in the low temperature
region, thus retained in the expansion for the crossection only the first
corresponding terms. This leads to incorrect behavior of the rate
coefficient at high temperatures $T>100K$, if compared with those for the H-
negative ion. Namely, it decreases and abruptly drops at $T_{k}=1500K$,
which approximately corresponds to redshift Z=500. For this reason, we need
to calculate this rate coefficient for a wider range of temperatures, from $%
T_{k}=100K$ to $3000K$. We have calculated this rate coefficient in our
paper \citep{b1} by using the principle of detailed balance and crossection
for the inverse process - photodetachment (see above), as suggested by %
\citet{b23}. In this way, we obtain a rate coefficient which has the correct
behavior with the kinetic temperature (approximately constant), like those
for the hydrogen radiative attachment ($H+e^{-}=H^{-}+\gamma $). It can be
approximately described by the following expression:

\begin{equation}
R_{ra}=1.53\cdot 10^{-15}(\frac{T_{r}}{300})^{-0.03}, [cm^{3}s^{-1}],
\end{equation}

which has been adopted in our model.

\subsection{Associative detachment}

There are two process of associative detachment of interest. The first one is

\begin{equation}
C^{-}+H=CH+e^{-},
\end{equation}

and the second is

\begin{equation}
H^{-}+C=CH+e^{-}.
\end{equation}

These were adopted in our calculations with the rate constant $5\cdot
10^{-10}$ and $10^{-9}$ $[cm^{3}s^{-1}]$ respectively from the UMIST data
base.

\subsection{Other important reactions}

Beside those mentioned above, we revisited and incorporated another process
which was not taken into account in our previous calculation \citep{b18}. As
it was mentioned in that paper, the main channel for the CH molecule
formation (destruction) are the neutral-neutral reactions $C(H_{2},H)CH$ and 
$CH(H,H_{2})C$. The most recent review and data on these processes was
presented by \citet{b8} (see also references therein).

As discussed by \citet{b18} one of the most important channels for the CH
molecule formation, is the $C+H_{2}=CH+H$ reaction. It was investigated in
detail in \citet{b7}, where the crossection for this process was proposed,
and reported to be in excellent agreement with the experimental data. The
rate coefficient for this reaction is given by the following expression

\begin{equation}
R_{1}=1.16\cdot 10^{-9}(\frac{T_{r}}{300})^{0.5}\exp (\frac{-14100}{T_{r}}),
[cm^{3}s^{-1}],
\end{equation}

The inverse process $CH+H=C+H_{2}$\ is the main channel of destruction for
the CH molecule. The most recent theoretical data on crossections and rate
coefficients were suggested by \citet{b8}, where a comparison with the
experimental data was also made (see references therein). The most recent
experimental data on this rate constant increases from $1.3\cdot 10^{-11}$ $%
cm^{3}s^{-1}$ at $T_{c}=300K$ to $1.6\cdot 10^{-10} $ $cm^{3}s^{-1}$\ at $%
T_{c}=2000K$. This fact suggests a small barrier of about $0.05eV$ \citep{b8}%
. The rate constant for this reaction, adopted in our calculations can be
represented by the following expression

\begin{equation}
R_{2}=9.0\cdot 10^{-11}(\frac{T_{r}}{300})^{0.3}\exp (\frac{-580}{T_{r}}%
)[cm^{3}s^{-1}],
\end{equation}

which is in good agreement with experimental results for the rate constant
of this process. It should be noted here that in the UMIST database for
Astrochemistry, a different expression for this rate constant is adopted,
but that it presents a discrepancy of almost an order of magnitude with the
experimental results, as well as with the theoretical one for a temperature
of 300K, and for these reasons it was unacceptable to use such rate for
these calculations.

\subsection{Details of the calculation and the abundance of CH}

The calculations of the molecular abundance were carried out for a wide
range of redshifts for the following species involved in the CH molecule
formation: $H$, $H^{+}$, $H_{2}^{+}$, $H_{2}$, $H^{-}$, $e^{-}$, $H_{3}^{+}$%
, $He$, $He^{+}$, $HeH^{+}$, $C$, $CH$, $CH_{2}$, $CH_{3}$, $CH_{4}$, $CH_{5}
$, $C^{+}$, $CH^{+}$, $CH_{2}^{+}$, $CH_{3}^{+}$, $CH_{4}^{+}$, $CH_{5}^{+}$%
, $C_{2}$, $C_{2}^{+}$, $C_{2}H$, $C_{2}H^{+}$, $C_{2}H_{2}$, $C_{2}H_{2}^{+}
$, $C_{2}H_{3}^{+}$, and $C^{-}$ . Other species are negligible in the
formation processes of the CH molecule because of their small abundance. In
our calculations we use the hydrogen and helium abundance according to the
Standard BBN and did not take into consideration deuterium and litium, due
to their insignificant role in the CH molecule production. We include helium
because of the importance of the $HeH^{+}$\ molecular ion in the production
of $H_{2}$\ by the chain of reactions $HeH^{+}+H=H_{2}^{+}+He$ ; $%
H_{2}^{+}+H=H^{+}+H_{2}$. The BBN carbon abundance was adopted from the
paper of \citet{b27} to be equal $[C]/[H]=10^{-9}$. It is an estimation
which can easily be changed to other abundance ratios, because it appears
linearly in the equations.

The results of the calculation are shown in Fig. 1 for the range of redshift
from $z=400$ and up to $z=10$. The relative abundance of the $CH$ molecule
is shown by a solid line and all the others by dashed lines. The $H$ and $He$
abundance are not shown in Fig. 1, since they are approximately constant,
and in this way it is possible choose a more convenient scale to see the
curves for $CH$, $CH^{+}$, etc. in more detail.

\begin{figure}[tbp]
\vspace{7cm} \includegraphics{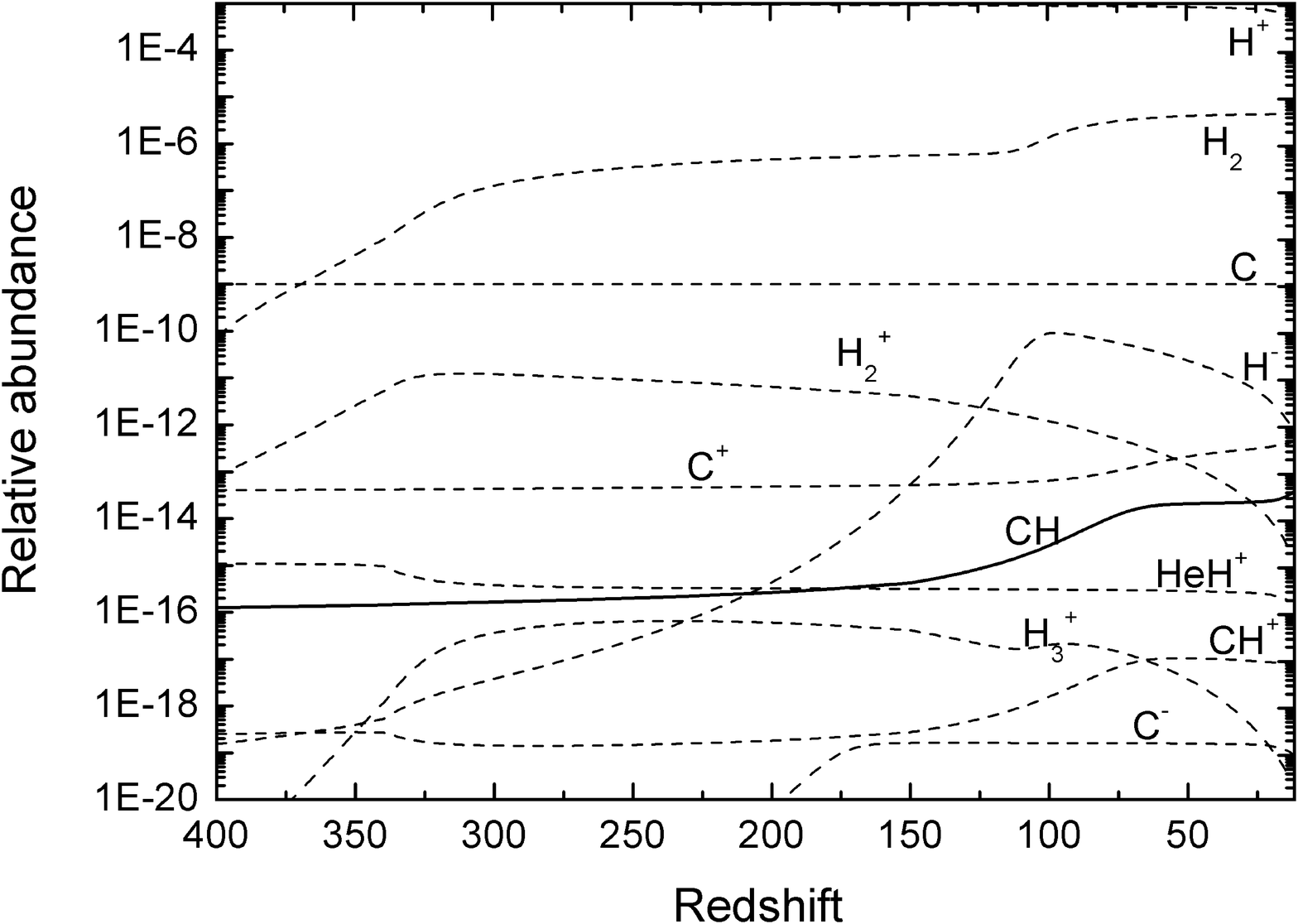} 
\caption{Fractional abundances of the chemical species involved in the CH
molecule formation as a function of redshift for an homogeneous primordial
medium in concordance with a $\Lambda $CDM cosmological model.}
\end{figure}

One can readily see in Fig. 1, that the relative abundance ratio $[CH]/[H]$
at low redshifts can be as high as a fraction of $10^{-13}$\ in the case of
an homogeneous medium. But naturally, inside a protoobject, the gas density
is expected to be higher, and so the chemistry must be faster and the
relative abundance of the $CH$ molecule will appear somewhere in the range
between $10^{-13}$\ and $10^{-9}$. To finalize, it should be noted that the
so called "minimum model" of $CH$ chemistry, suggested in \citet{b18} is
incomplete at low temperatures, and more species (such as $CH_{2}$, $CH_{3}$%
, $C^{+}$, $CH^{+}$, $CH_{2}^{+}$, $CH_{3}^{+}$, and $C^{-}$ ) must be
included in the calculations.

\section{Optical depth and observational estimates}

It is well known that the largest resonant line opacities in homogeneous
mass perturbations are produced when they are at their maximum expansion or
turn-around epoch (linear stage of their evolution), when all parts of the
protocloud participate in the line formation process.

Let us estimate the detectability of line features of the primordial $CH$
molecule in protostructures still in their linear or about-linear evolution
regime, i.e. before gravitational collapse and first star formation occurs.
Here we calculate the CH resonant line opacities only at the turnaround
epoch for CDM high-density ($6-\sigma $) mass perturbations. The interaction
of CMBR photons with the primordial molecules inside these protostructures
will produce spectral - spatial signatures in the CMBR spectrum, containing
information on the physical conditions of the gas prior to the formation of
the first stars and minigalaxies in the universe.

The optical depths of primordial protoclouds in the rotational lines of CH
molecule were calculated using the spherical top-hat approach (see for
example \citet{b25,b32}, to obtain the epoch, density, and size of the CDM
mass perturbations at their maximum expansion, for the $\Lambda $CDM
cosmology. We have discussed the top-hat approach in our previous paper (see %
\citet{b24}), and the parameters of the first structures listed in Table 1
and Table 2 of that paper will be used here. Table 2, in particular,
presents the parameters we need for $6\sigma $ halos such as the turn-around
redshift, size of protoclouds, density etc.

The optical depth for a protocloud of size $L$ is given by: 
\begin{equation}
\tau _{\nu }\left( L\right) =\int_{0}^{L}\alpha _{\nu }dx,  \label{21}
\end{equation}%
where the integration is carried out over the line of sight and $\alpha
_{\nu }\left( x\right) $\ is the absorption coefficient given by the
following expression: 
\begin{equation}
\alpha _{\nu }=\frac{\lambda ^{3}\left( 2J^{\prime }+1\right) }{8\pi \left(
2J+1\right) V_{T}}x_{CH}n_{J}n_{tot}A_{J^{^{\prime }}J}\left( 1-e^{\frac{%
-h\nu }{\kappa T_{r}}}\right) .  \label{tau2}
\end{equation}%
Here $\lambda $\ is the wavelength, $J$\ is the rotational quantum number, $%
x_{CH}$\ is the relative abundance of the $CH$ molecule shown in fig. 1, $%
n_{J}$\ is the population of $J$-th rotational level at epoch $z$, and $%
A_{J^{^{\prime }}J}$\ is the Einstein coefficient.

By using these expressions, we calculate the $CH$ line optical depths
corresponding to protohalos of several masses at their turnaround redshifts.
The result is shown in Fig. 2 for three ground rotational line transitions
3/2($^{2}\Pi _{1/2}$)-1/2($^{2}\Pi _{1/2}$), 5/2($^{2}\Pi _{1/2}$)-3/2($%
^{2}\Pi _{1/2}$) and 3/2($^{2}\Pi _{3/2}$)-1/2($^{2}\Pi _{1/2}$) for the $%
6\sigma $ first halos.

As we can see, the values of the optical depth for the $CH$ molecular lines
are much larger than those reported for the HD molecule \citep{b24} and
reach values $10^{-4}$ for redshift $z=20 $.


\begin{figure}[tbp]
\vspace{7cm} \includegraphics{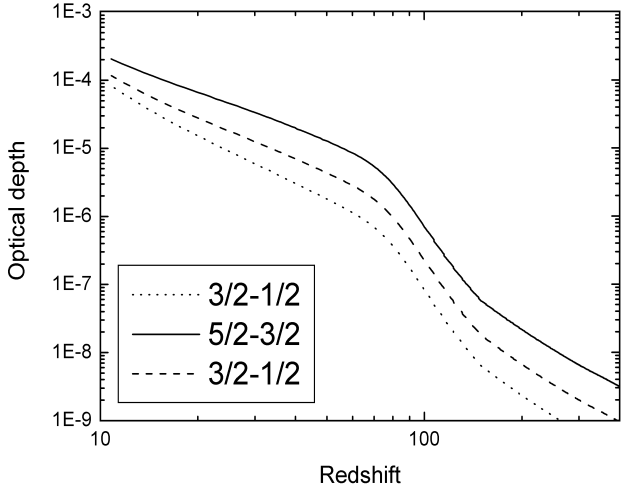} 
\caption{Optical depths in the first, second and third rotational
transitions of CH molecule in $6\protect\sigma $ $\Lambda $CDM protohalos of
different masses reaching their maximum expansion at the redshifts shown in
the abscissa. Dotted and solid lines correspond to the $^{2}\Pi _{1/2}$
state, and the dashed line is for 3/2($^{2}\Pi _{3/2}$) - 1/2($^{2}\Pi
_{1/2} $) rotational transition.}
\end{figure}

Let us now estimate some observational parameters for the first structures
in the $CH$ rotational lines. The temperature fluctuations of the CMBR due
to the resonance scattering of the CMBR photons by moving protoclouds is: %
\citep{b29}%
\begin{equation}
\frac{\Delta T}{T}=\frac{V_{p}}{c}\tau .  \label{deltaT}
\end{equation}%
Here $V_{p}=V_{p}(z)$ is the peculiar velocity of the protocloud at epoch $z$
with respect to the CMBR, $c$ is the speed of light, and $\tau $\ is the
corresponding optical depth of the protocloud at epoch $z_{ta}$ .

The linear theory of gravitational instability shows that the peculiar
velocity of every mass element grows with the expansion factor as $%
V_{p}(z)\propto \dot{D}(z)/(1+z)$. An accurate approximation to this
expression for a flat universe with a cosmological constant is suggested in
the papers \citet{b13} and \citet{b3}. The value of the peculiar velocity at
the present epoch $V_{p}(z=0)\approx 650$ km/s, adopted in our calculation,
was taken from \citet{b17}, \citet{b9,b10,b35}. By using this expression and
the $CH$ optical depths shown at Fig. 2, one can estimate with eq. (\ref%
{deltaT}) the temperature fluctuation of the CMBR secondary anisotropies. In
our case for the redshift $z=20$, for the case of the $6\sigma $ protohalos,
we obtain that $\Delta T/T\approx \ 2\cdot 10^{-7}$ in the 5/2-3/2
rotational transition at the laboratory wavelength $\lambda _{0}=181\mu m$,
whereas for the case of collapsed halos this value can be higher.

Let us now estimate the integration time required for the detection of the
secondary anisotropies with modern telescopes (such as ALMA, CARMA or
LMT/GTM). The observational time $\Delta t$\ can be estimated from the
equation 
\begin{equation}
\Delta T=\frac{T_{n}}{\sqrt{\Delta \nu \Delta t}},  \label{25}
\end{equation}%
where $\Delta \nu $\ is the bandwidth, $T_{n}$ is the noise temperature and $%
\Delta T$\ is the amplitude of the temperature fluctuations calculated
above. For $\Delta \nu $ we adopt an estimation $\Delta \nu /\nu \approx 0.1$
and $T_{n}\approx 50$ K. Thus, for the $6\sigma $ protohalos, the
integration time required for an observable signal of the rotational
transitions of CH molecules is estimated to be $\Delta t\approx 7\cdot
10^{5}s=200\ $\ hours, which can be divided into 10 - 20 sets of
observations near the north (south) pole regions. It should be stressed here
that for the case of collapsed gas inside virialized $3\sigma $ halos,
previous to star formation triggering (see for detail the discussion in %
\citet{b24}), the integration time needed for the detection of the
rotational lines of the primordial CH molecule, can be reduced to up to a
few hours.

To conclude this part, the angular size of the objects must be mentioned.
The first halos discussed in the present paper are actually the same
considered in our previous paper \citep{b24}, for which the signatures of
the HD primordial molecule were investigated. For this reason we do not need
any special consideration here and can refer to Fig. 3 in that paper. It
should be noted that the angular size for all protoclouds reaching their
maximum expansion in the redshift range of $20<z<40$ (this corresponds
roughly to a mass range of $10^{5}<M/\mbox{M$_{\odot}$}<10^{9}$ and $%
10^{9}<M/\mbox{M$_{\odot}$}<10^{11} $ for the $3\sigma $ and $6\sigma $
cases, respectively) appears within the observable region of ALMA and CARMA.

\section{Conclusions}

In the present paper we present for the first time the calculation of the
kinetics of the primordial $CH$ molecule and the observational estimates for
protohalos during the first structure formation epoch, which can possibly be
observed in the $CH$ molecule rotational lines. The signature of the first
structure formation epoch is actually the Spectral-Spatial Fluctuations in
the CMBR spectrum due to the elastic resonant scattering of the CMBR by
primordial $CH$ molecules in its rotational structure. The optical depths in
the pure rotational structure of the molecule were calculated for the $%
6\sigma $ halos within the framework of a $\Lambda $CDM universe.

For our calculation, we adopt the relative abundance of primordial carbon $%
[C]/[H]=10^{-9}$ taken from the paper of \citet{b27}. By taking into account
the fact that the relative abundance of the $CH$ primordial molecule depends
linearly on the $[C]/[H]$ value, one can easily reestimate the results for
any particular case of NBBN. The main results of this work can be summarized
as follows:

1) We calculate for the first time the chemistry of primordial carbon in the
case of Nonstandard BBN. This was done for an homogeneous primordial gas.

2) It is shown that the most abundant of the carbon-beared molecules is CH,
whereas others ( such as $CH_{2}$, $CH_{3}$, $CH_{4}$, $CH_{5}$, $CH_{2}^{+}$%
, $CH_{3}^{+}$, $CH_{4}^{+}$, $CH_{5}^{+}$, $C_{2}$, $C_{2}^{+}$, $C_{2}H$, $%
C_{2}H^{+}$, $C_{2}H_{2}$, $C_{2}H_{2}^{+}$ ) only appear in negligible
amounts. This difference, if compared with the interstellar medium, can be
explained in terms of the rather low abundance of primordial gas and carbon.
It must be stressed that in the presence of inhomogeneites, the CH molecule
abundance increases, so that in this sense, these calculations can be
considered as lower limit estimations.

3) Within the framework of the $\Lambda $CDM universe we calculate the
optical depth for pure rotational lines of the primordial $CH$ molecule for
first structures formed during the Dark Age epoch. At redshifts $z=20$, for
the $6\sigma $ model, the calculated optical depth is $\tau _{\nu }\approx
10^{-4}$, in two ground and one exited rotational transitions. The
frequencies of observation for these three lines are $\nu \approx
25,80,96GHz $ that correspond respectively, to the redshifted lines of the
rotational transitions $(J\prime J)=(3/2-1/2)$; $(5/2-3/2)$; $(3/2-1/2)$ of
the $CH$ molecule (see Fig. 2). We would like to stress here, that the
presence of tree observable lines with not too different optical depths,
would permit to identify the observed object as a first halo with a redshift 
$20-30$.

4) For testing NBBN models, the $CH$ molecule is a more efficient tool in
searches for signatures of the first structure formation epoch, than the $HD$
molecule, discussed in our previous paper \citep{b24}. Furthermore, the CH
rotational lines (even if not detected) would allow us to discriminate
between the NBBN models.\ The optical depth of first halos in rotational
lines of $CH$ in the case of NBBN model discussed in the paper of \citet{b27}%
, is larger than that for $HD$ by a factor of $10^{2}$.

5) In the case of the collapsed $3\sigma -6\sigma $\ halos (see \citet{b24}%
), the optical depth must increase by $100-200$ times and the observational
time would be reduced to a fraction of one hour for the NBBN model of %
\citet{b27}, and probably even the model of IBBN suggested by \citet{b15}
with $[C]/[H]=3\cdot 10^{-13}$ could be detected, by taking into account
that the enhancement of the CH fractional abundance during the collapse of
the protocloud, leads to detectable fluctuations in the CMBR temperature.
For this reason it is of great importance to do detail investigation of the
non-linear stages of collapse at the first structure formation epochs.

6) As it was mentioned above, the first stars (PopIII) possibly appeared at $%
Z\approx 20$, and significantly enriched the environments with heavy
elements. In this case the relative abundance of $^{12}C$ at $Z=19$ due to
feedback can reach $[C]/[H]=10^{-4}$ \citep{b4}, an abundance that is
expected to be detected with modern radiotelescopes.

To conclude, we would like to stress again that the optical depths in the
rotational lines of the CH molecule from the first structures depend
strongly on the models of NBBN and structure formation, and that in some
cases these lines might be observable today, whereas for others it is not
yet possible. In any case, this work emphasizes the need to search for
signatures of these objects in order to find them or, at least, to put upper
limits to constrain scenarios of formation and NBBN models.

\section{Acknowledgments}

This work was supported in part by PROMEP proyecto 12507 0703027 and PAPIIT
IN107706-3

\end{document}